\documentclass[twocolumn,amsmath,amssymb,floatfix,nofootinbib]{revtex4}

\usepackage{epsfig}
\usepackage{amsmath,amssymb}
\usepackage[all]{xy}

\newcommand{\beq}{\begin{equation}}
\newcommand{\eeq}{\end{equation}}
\newcommand{\lsi}{\,\raisebox{-0.13cm}{$\stackrel{\textstyle<}
{\textstyle\sim}$}\,}
\newcommand{\gsi}{\,\raisebox{-0.13cm}{$\stackrel{\textstyle> 
}
{\textstyle\sim}$}\,}

\begin{document}

\title {\bf A Window in the Dark Matter Exclusion Limits}
\author{Gabrijela Zaharijas and Glennys R. Farrar}
\affiliation{\it Center for Cosmology and Particle Physics\\  New York University, NY, NY 10003,USA}

\begin{abstract}

We consider the cross section limits for light dark matter candidates ($m=0.4$ to 10 GeV). We calculate the interaction of dark matter in the crust above underground dark matter detectors and find that in the intermediate cross section range, the energy loss of dark matter is sufficient to fall below the energy threshold of current underground experiments. This implies the existence of a window in the dark matter exclusion limits in the micro-barn range.

\end{abstract}

\maketitle

\vspace{4pt}

\section{Introduction}
The evidence for the existence of dark matter (DM) is manifold; in particular it comes from measurements of galactic rotational curves, our understanding of structure formation and from the CMB spectrum. Recent WMAP observations determined $\Omega _m h^2=0.135^{+.008} _{-.009}$ and the baryonic fraction of matter $\Omega _{b}\Omega ^{-1} _{m}=0.17\pm 0.01$ \cite{wmap}. 

In this paper we explore the possibility that the DM mass is in a range close to hadronic masses, $0.4 \lsi m \lsi 10$ GeV. Masses below $\sim 0.4$ GeV are below the threshold of direct detection DM experiments and are therefore unconstrained, with the exception of axion searches. The region of mass $m \gsi 10$ GeV is well explored today, up to TeV range, from strong ($\sigma \sim 10$ mb) to weak ($\sigma \sim 10^{-6}$pb) cross sections, and a new generation of experiments reaching $\sigma \sim 10^{-7,8}$pb is planned for the near future. 

The mass range $m\lsi 10$ GeV has not yet been explored carefully. Several dark matter underground experiments have sufficiently low mass threshold today: the CRESST \cite{cresst}, DAMA \cite{dama}, IGEX \cite{igex}, COSME \cite{cosme} and ELEGANT \cite{elegant} experiments. Except for DAMA, these experiments have published upper limits on the cross section assuming it is weak, but have not addressed the case of stronger cross sections,\footnote{DAMA did publish strong cross section limits in \cite{damastrong}, but they were based on a dedicated experiment which had a higher mass threshold $m\gsi 8$ GeV.} where the approach for extracting the cross section limits is substantially different, as we explain below. Also, recent data from an X-ray experiment XQC, \cite{XQC} proved to be valuable in constraining low mass DM, but limits based on the final data have not yet been derived. Since XQC is a space\-- based experiment it is especially suitable for exploring the higher cross section range. In \cite{SS} it was shown that in the low mass range the XQC experiment rules out Strongly Interacting DM (SIMPs, \cite{SpergelSteinhardt}). Dark matter with low masses and 'intermediate' cross sections, several orders of magnitude smaller than normal hadronic cross sections, remains to be fully analyzed and that is the focus of this work. We will abbreviate DM with intermediate cross section on nucleons as DMIC.       

One of the motivations for exploring the low mass range is the model with baryon and antibaryonic dark matter, which offers a simultaneous solution for the baryon asymmetry and dark matter problems, introduced in \cite{f,fz}. There, it has also been shown that dark matter which carries baryon number $b_{X}$ needs to be lighter than $4.5~|b_{X}|$ GeV in order for the Universe to have zero net baryon number. In such a scenario it can be natural for the DM elastic cross to be in the intermediate range, \cite{f,fz}. 

Early limits from DMIC direct detection experiments can be found in the paper \cite{RRS} by Rich, Rocchia and Spiro in which they reported results from a 1987 balloon experiment. Starkman et al. \cite {SG} reviewed DM constraints down to a mass of 1 GeV as of 1990. Wandelt et al. \cite{SS} added constraints based on preliminary data from the more recent XQC sounding rocket experiment. The above constraints are discussed and updated further in the text. In previous works on this topic the effect of the halo particle's velocity distribution on the cross section limit was not explored. Since the only detectable light particles are those in the exponential tail of the velocity distribution, the limits on light DM are sensitive to the parameters in the velocity distribution, in particular to the value of the escape velocity cutoff. We investigate this sensitivity in the standard Isothermal Sphere model, where the DM velocity distribution function is given by a truncated Maxwell-Boltzmann distribution. We also consider the spin-independent and spin-dependent interaction cases separately. Except in Section \ref{fraction}, we assume a single type of DM particle.

\section{Direct Dark Matter Detection} \label{directdet}

The basic principle of DM detection in underground experiments is to measure the nuclear recoil in elastic collisions, see for example \cite{Lewin}. The interaction of a DM particle of mass $m\lsi 10$ GeV, produces a recoil of a nucleus of 20 keV or less. The recoil energy (which grows with DM mass as in (2) below) can be measured using various techniques. For example, in the CRESST experiment, Al$_2$O$_3$ crystal was used and the temperature rise from the excitation of phonons due to nucleus \-- DM scattering, was measured to extract the recoil energy. 

For a given velocity distribution $f({\vec v})$, the differential rate per unit recoil energy $E_R$ in (kg day keV)$^{-1}$ in the detector can be expressed as
\beq 
\label{cr}
\frac {dR}{dE_R} = N_T~n_{X} \int _{v_{min}} ^{v_{esc}} ~d{\vec v}~|{\vec v}|~f({\vec v})~g({\vec v}) \frac{d \sigma_{XA}}{dE_R}, 
\eeq
where $n_{X}$ is the number density of DM particles, $N_T$ is the number of target nuclei per kg of target, $\sigma _{XA}$ is the energy dependent scattering cross section of DM on a nucleus with mass number A, $g({\vec v})$ is the probability that a particle with velocity $v$ deposits an energy above the threshold $E_{TH}$ in the detector, and $v_{min}$ is the minimum speed the DM particle can have and produce an energy deposit above the threshold. 
The recoil energy of the nucleus is given by 
\beq \label{er}
E_{R}=\frac {4m_A~m_{X}}{(m_A+m_{X})^2} (\frac {1}{2} m_X v^2 _{X}) \left( \frac {1-\cos\theta _{CM}}{2}\right)
\eeq
where $\theta _{CM}$ is the scattering angle in the DM-nucleus center of mass frame. We will assume isotropic scattering as is expected at low energies. 
So, for instance, for $A=16$, $m=1$ GeV and an energy threshold of 600 eV, the minimal DM velocity to produce a detectable recoil is $v_{min}=680$ km/s, in the extreme tail of the DM velocity distribution.

In order to compare cross section limits from different targets we will normalize them to the proton-DM cross section, $\sigma _{Xp}$. For the simplest case of interactions which are independent of spin and the same for protons and neutrons, the low energy scattering amplitude from a nucleus with mass number A is a coherent sum of A single nucleon scattering amplitudes. The matrix element squared therefore scales with size of nucleus as $\sim A^2$. In addition the kinematic factor in the cross section depends on the mass of the participants in such a way \cite{witten,Lewin} that 
\beq \label{sigSI}
\frac {\sigma^{SI} _{XA}}{\sigma^{SI} _{Xp}}=\left( \frac {\mu (A)}{\mu(p)}\right) ^2~A^2
\eeq
where $\mu (A)$ is the reduced mass of the DM-nucleus system, and $\mu (p)$ is the reduced mass for the proton-DM system.  
At higher momentum transfer $q^2=2m_NE_R$ the scattering amplitudes no longer add in phase, and the total cross section $\sigma _{XA} (q)$ becomes smaller proportionally to the form factor $F^2(q^2)$, $\sigma _{XA} (q)=\sigma _0 F^2(q^2)$.   

We take this change in the cross section into account when we deal with higher mass ($m\gsi 10$ GeV) dark matter; for smaller masses the effect is negligible. We adopt the form factor $F(q^2)=\exp\left(-1/10(qR)^2\right)$ with $R=1.2 A^{1/2}$ fm, used also in \cite{formfactor,formfactor2}. The simple exponential function is suffitiently accurate for our purposes and easy to implement using the Monte Carlo method to sample momentum transfer $q$, from its distribution given by the form factor. The procedure is described in more detail in Appendix B. 

For spin dependent interactions the scattering amplitude changes sign with the spin orientation. Paired nucleons therefore contribute zero to the scattering amplitude and only nuclei with unpaired nucleon spins are sensitive to spin dependent interactions. Due to the effect of coherence, the spin independent interaction is usually dominant, depending on the mass of the exchanged particle \cite{kamionkowski}. Therefore, the spin dependent cross section limit is of interest mainly if the spin independent interaction is missing, as is the case, for example, with massive majorana neutrinos. Another example of DM with such properties is photino dark matter, see \cite{witten}, in the case when there is no mixing of left- and right- handed scalar quarks. The amplitude for DM-nucleus spin dependent interaction in the case of spin 1/2 DM, in the nonrelativistic limit, is proportional to \cite{witten, engel-vogel} 
\beq
{\cal M}\sim <N|{\vec J}|N>\cdot {\vec s}_{X}
\eeq
where ${\vec J}$ is the total angular momentum of the nucleus, $|N>$ are nuclear states and ${\vec s}_{X}$ is the spin of the DM particle. 
In the case of scalar DM the amplitude is 
\beq
{\cal M}\sim <N|{\vec J}|N>\cdot \left( {\vec q} \times {\vec q}' \right)
\eeq
where ${\vec q}$ and ${\vec q}'$ are the initial and final momenta of the scattering DM particle. Thus the cross section for this interaction is proportional to the fourth power of the ratio $q/M$, of DM momentum to the mass of the target which enters through the normalization of the wavefunction. Therefore the spin dependent part of the interaction for scalar DM is negligible when compared to the spin independent part. 

We adopt the standard spin-dependent cross section parametrization \cite{Lewin}
\beq 
\sigma _{XA}\sim \mu (A) ^2~[\lambda ^2J(J+1)]_A C^2 _{XA} 
\eeq 
where $\lambda$ is a parameter proportional to the spin, orbital and total angular momenta of the unpaired nucleon. The factor $C$ is related to the quark spin content of the nucleon, $C=\sum T^q _3 \Delta_q,~q=u,d,s$, where $T^{u,d,s} _3$ is the charge of the quark type $q$ and $\Delta_q$ is the fraction of nucleon spin contributed by quark species $q$.
The nuclear cross section normalized to the nucleon cross section is
\beq \label{sigSD}
\frac {\sigma^{SD} _{XA}}{\sigma^{SD} _{Xp}}=\left( \frac {\mu (A)}{\mu(p)}\right) ^2~\frac {[\lambda ^2J(J+1)]_A}{[\lambda ^2J(J+1)]_p}\left( \frac {C_{XA}}{C_{Xp}}\right)^2 .
\eeq 
The values of proton and neutron $C$ factors, $C_{Xp},~C_{Xn}$ vary substantially depending on the model. For targets of the same type \-- odd-n (Si, Ge) or odd-p (Al, Na, I) nuclei \-- this model dependence conveniently cancels. The comparison of cross sections with targets of  different types involves the $C_{Xp}/C_{Xn}$ ratio. This ratio was thought to have the value  $\sim 2$  for any neutralino, based on the older European Muon Collaboration (EMC) measurements, but the new EMC results imply a ratio which is close to one for pure higgsino, and is $\gsi$ 10 otherwise. (The biggest value for the ratio is $C_p/C_n\sim 500$, for bino.) We normalize our spin dependent results to the proton cross section $\sigma _{Xp}$  using  $C_{Xp}/C_{Xn}=1$ for definiteness below. 

In this paper we assume that the DM halo velocity distribution is given by a truncated Maxwell-Boltzmann distribution in the galactic reference frame, as in the Isothermal Sphere Halo model \cite{Binney}. We direct the ${\hat z}$ axis of the Earth's frame in the direction of the Local Standard of Rest (LSR) motion. \footnote{The Local Standard of Rest used here is the dynamical LSR, which is a system moving in a circular orbit around the center of Milky Way Galaxy at the Sun's distance.} The DM velocity distribution, in the Earth's frame, is given by
\beq
\label{veldistE}
f(v_{z},{\vec v}_{\perp})=N \exp \left[-\frac{(v_{z}-v^{t} _E)^2+{\vec v}_{\perp}^2}{v^2 _c}\right].
\eeq
Here $v_c$ is the local circular velocity and it is equal to $\sqrt{2}$ times the radial velocity dispersion in the isothermal sphere model; ${\vec v}_E$ is the velocity of the Earth in the Galactic reference frame. Throughout, superscript ``t'' indicates a tangential component. This neglects the Earth's motion in the radial direction which is small. The velocities $v_z$ and ${\vec v}_{\perp}$ are truncated according to $\sqrt {v^2 _z+{\vec v}_{\perp}^2}\lsi v_{esc}$, where $v_{esc}$ is the escape velocity discussed below. 

The model above is the simplest and the most commonly used model which describes a self-gravitating gas of collisionless particles in thermal equilibrium. On the other hand numerical simulations produce galaxy halos which are triaxial and anisotropic and may also be locally clumped depending on the particular merger history (see \cite{annegreen} for a review). This indicates that the standard spherical isotropic model may not be a good approximation to the local halo distribution. Here we aim to extract the allowed DM window using the simplest halo model, but with attention to the sensitivity of the limit to poorly determined parameters of the model. The effects of the more detailed halo shape may be explored in a further work. 

We ignore here the difference between the DM velocity distribution on the Earth, deep in the potential well of the solar system, and the DM velocity distribution in free space. This is a common assumption justified by Gould in \cite{gould} as a consequence of Liouville's theorem. Recently Edsjo et al. \cite{edsjo} showed that the realistic DM velocity distribution differs from the free space distribution, but only for velocities $v\lsi 50$ km/s. Therefore, the free space distribution is a good approximation for our analysis, since for light dark matter the dominant contribution to the signal comes from high velocity part of the distribution.    

The velocity of the Earth in the Galactic reference frame is given by
\beq
{\vec v}_E={\vec v}_{LSR}+{\vec v}_{S}+{\vec v}_{E,orb},
\eeq
where ${\vec v}_{LSR}$ is the velocity of the local standard of rest LSR: it moves with local circular speed in tangential direction $v^t _{LSR}=v_c$, toward $l=90^o$, $b=0^o$, where $l$ and $b$ are galactic longitude and latitude.  The velocity of the Sun with respect to the LSR is ${\vec v}_{S}= 16.6$ km/s and its direction is $l=53^o$, $b=25^o$ in galactic coordinates. $v_{E,orb}=30$ km/s is the maximal velocity of the Earth on its orbit around the Sun.

The magnitude of $v^t _{LSR}$ has a considerable uncertainty. We adopt the conservative range $v_c=(220\pm 50)$ km/s which relies on purely dynamical observations \cite{kochanek}. Measurements based on the proper motion of nearby stars give a similar central value with smaller error bars, for example $v_c (R_0)=(218\pm 15)$ km/s, from Cepheids and $v_c (R_0)=(241\pm 17)$ km/s, from SgrA$^*$ (see \cite{annegreen2} and references therein). The choice $v_c=(220\pm 50)$ km/s is consistent with the DAMA group analysis in \cite{damav} where they extracted the dependence of their cross section limits on the uncertainty in the Maxwellian velocity distribution parameters.

Projecting the Earth's velocity on the tangential direction ($l=90^o$, $b=0^o$) we get 
\beq
v^{t} _E=v_c+v^{t} _S + v^{t} _{E,orb}~\cos [\omega (t-t_0)]
\eeq
where $v^{t} _S=12~{\rm km/s}$; $v^{t} _E= 30~\cos \gamma~{\rm km/s}$ where $\cos \gamma=1/2$ is the cosine of the angle of the inclination of the plane of the ecliptic, $\omega =2\pi/365$ day$^{-1}$ and $t_0$ is June 2nd, the day in the year when the velocity of the Earth is the highest along the LSR direction of motion. In the course of the year $\cos [\omega (t-t_0)]$ changes between $\pm ~1$, and the orbital velocity of the Earth ranges $\pm 15$ km/s.
Taking all of the uncertainties and annual variations into account, the tangential velocity of the Earth with respect to the Galactic center falls in the range $v^{t} _{E}=(167~{\rm to}~307)~{\rm km/s}$.
 
The other parameter in the velocity distribution with high uncertainty is the escape velocity, $v_{esc}=(450~{\rm to}~650)$ km/s \cite{vesc}. We will do our analysis with the standard choice of velocity distribution function parameters, 
\beq \label{parameters}
v^t _{E}=230~ {\rm km/s},~~ v_c=220~ {\rm km/s},~~v_{esc}=650~ {\rm km/s},
\eeq 
and with the values of $v_E$ and $v_{esc}$ from their allowed range, which give the lowest count in the detector and are therefore most conservative:
\beq \label{range}
v^t _{E}=170~ {\rm km/s},~~v_c=170~ {\rm km/s},~~v_{esc}=450~ {\rm km/s}.
\eeq
For experiments performed in a short time interval we take the value of $v^{t} _{E,orb}~\cos [\omega (t-t_0)]$ which corresponds to the particular date of the experiment, and the lowest value of $v^t _{E}$ allowed by the uncertainties in the value of $v_c$.
 
Another effect is important in the detection of moderately interacting particles. Since particles loose energy in the crust rapidly (mean free path is of the order of 100 m) only those particles which come to the detector from $2\pi$ solid angle above it can reach the detector with sufficient energy. Since the velocity distribution of the particles arriving to the detector from above depends on the detector's position on Earth with respect to the direction of LSR motion, the detected rate for a light DMIC particle will vary with the daily change of position of the detector. This can be a powerful signal.

\section{XQC Experiment}

For light, moderately interacting dark matter the XQC experiment places the most stringent constraints in the higher cross section range.
The XQC experiment was designed for high spectral resolution observation of diffuse X-ray background in the $60-1000$ eV range. The Si detector consisted of an array of 36 1 mm$^2$ microcalorimeters.  Each microcalorimeter had a 7000 Angstrom layer of HgTe X-ray absorber. Both the HgTe and the Si layers were sensitive to the detection. The experiment was performed in a 100 s flight in March, and therefore the Earth's velocity $v^t _{E}$ falls in the 200 to 300 km/s range. The experiment was sensitive to energy deposit in the energy range $25-1000$ eV. For energy deposits below 25 eV the efficiency of the detector drops off rapidly. For energy deposits above about 70 eV the background of X-rays increases, so XQC adopted the range 25-60 eV for extraction of DM limits, and we will do the same. This translates into a conservative mass threshold for the XQC experiment of $0.4$ GeV, obtained with $v_{esc}=450$ km/s and $v^t _E=200$ km/s, which is the lowest mass explored by direct DM detection apart from axion searches.

The relationship between the number of signal events in the detector $N_S$ and the scattering cross section $\sigma_{XA}$ of DM particles on nuclei is the following
\begin{eqnarray} \label{rate}
N_S &=& n_X~f~T~( N_{\rm Si}  <{\vec v}_{\rm Si}> \sigma_{\rm Si}\\
&+& N_{\rm Hg} ( <{\vec v}_{\rm Hg}> \sigma_{\rm Hg}+<{\vec v}_{\rm Te}> \sigma_{\rm Te} )), \nonumber
\end{eqnarray}
where $N_{\rm Si}$ and $N_{\rm Hg}$ are the numbers of Si and Hg (Te) nuclei in the detector, $n_{X}$ is the local number density of DM particles, $<\vec{v}_{\rm Si}>$, $<\vec{v}_{\rm Hg}>$ and $<\vec{v}_{\rm Te}>$ are the effective mean velocities of the DM particles on the Si and HgTe targets, $f$ is the efficiency of the detector, and $T=100$ s is the data-taking time. In this energy range, $f \approx 0.5$.
The standard value for the local DM energy density is $\rho _X=0.3$ GeV cm$^{-3}$. However, numerical simulations combined with observational constraints \cite{draco} indicate that the local DM energy density $\rho _X$ may have a lower value, $0.18 \lsi \rho _X/({\rm GeV}~{\rm cm}^{-3}) \lsi 0.3 $. In our calculations we use both the standard value $\rho _{X}=0.3$ GeV/cm$^3$, and the lower value suggested by the numerical simulations, $\rho _{X}=0.2$ GeV/cm$^3$. The cross sections $\sigma_{\rm Si}$, $\sigma_{\rm Hg}$, $\sigma_{\rm Te}$ are calculated using equations (\ref{sigSI}) and (\ref{sigSD}). In this section and the next we assume that DM has dominantly spin-independent cross section with ordinary matter. In Section \ref{SD} we consider the case of DM which has exclusively spin-dependent cross section or when both types of interaction are present with comparable strength.
 
XQC observed two events with energy deposit in the 25-60 eV range, and expected a background of 1.3 events. The equivalent 90\% cl upper limit on the number of signal events is therefore $N_S = 4.61$. This is obtained by interpolating between 4.91 for expected background = 1.0 event and 4.41 for expected background = 1.5 events, for 2 observed events using table IV in ref. \cite{feldman}.

We extract the cross section limits using our simulation. Because of the screening by the Earth we consider only particles coming from the $2\pi$ solid angle above the detector, for which $<{\hat n} \cdot {\vec v}>\leq 0$ and for them we simulate the interaction in the detector, for particles distributed according to (\ref{veldistE}). We take the direction of the LSR motion, ${\hat n}$ as the z axis. 

We choose the nucleus $i$ which the generated DM particle scatters from, using the relative probability for scattering from nucleus of type $i$, derived in Appendix A: 
\beq \label{probability}
P_i =\frac{\lambda _{eff}}{\lambda _i}=\frac {n_i\sigma _{XA_i}}{\sum  n_j\sigma _{XA_j}},
\eeq 
where $\lambda _i$ is the mean free path in a medium consisting of material with a mass number $A_i$: $\lambda _i=(n_i \sigma_{XA_i})^{-1}$. Here $n_i$ is the number density of target nuclei $i$ in the crust, $\sigma_{XA_i}$ is the scattering cross section of X on nucleus A$_i$ and the effective mean free path, $\lambda _{eff}$, is given as 
\beq \label{freepath}
\lambda _{eff}=\left( \sum \frac {1}{\lambda _i} \right) ^{-1}.
\eeq
In each scattering the DM particle loses energy according to (\ref{er}), and we assume isotropic scattering in the c.m. frame.

We determine the effective DM velocity $<{\vec v_A}>$ as
\beq \label{<v>}
<{\vec v_A}>=\frac {\sum ' v}{N_{tot}}
\eeq
where the sum is over the velocities of those DM particles which deposit energy in the range 25-60 eV, in a collision with a nucleus of type A, and $N_{tot}$ is the total number of generated DM particles. The result depends on the angle between the experimental look direction, and the motion of the Earth. 
The zenith direction above the place where the rocket was launched, $\hat n _{XQC}$, is toward  $b=+82^o,l= 75^o$. Thus the detector position angle compared to the direction of motion of the Earth through the Galaxy is 82$^o$. Only about $10\%$ of the collisions have an energy deposit in the correct range.
Putting this together gives the $90 \% $ confidence level XQC upper limit on the spin independent cross section for DMIC shown in Figures 1 and 2. The solid line limit is obtained using the most conservative set of parameters ($\rho =0.2$ GeV/cm$^3$, $v^t _E=200$ km/s, $v_{esc}=450$ km/s) and the dotted line is the limit obtained by using the standard parameter values in equation (\ref{parameters}); The upper boundary of the upper domain, $\sigma= (10^6$ to $10^8)$ mb is taken from \cite{SS}. 

When the dark matter mass is higher than 10 GeV, the form factor suppression of cross section is taken into account. We give the details of that calculation in the Appendix B.

In the next section we explain how the upper boundaries of the excluded region from the underground experiments shown in these figures are obtained. Also shown are the original limits from the balloon experiment of Rich et al. \cite{RRS} obtained using the ``standard'' choices for DM at the time of publishing (dashed line) as well as the limits obtained using the conservative values of parameters ($v^t _E=170$ km/s, since the experiment was performed in October, and $v_{esc}=450$ km/s). Fig. 2 zooms in on the allowed window in the $m\lsi 2.4$ GeV range.

\section{Underground Detection}

In this section we describe the derivation of the lower boundary of the DMIC window from the underground experiments.
This is the value of $\sigma _{Xp}$ above which the DM particles would essentially be stopped in the Earth's crust before reaching the detector. (More precisely, they would loose energy in the interactions in the crust and fall below the threshold of a given experiment.) To extract the limit on $\sigma _{Xp}$ we generate particles with the halo velocity distribution and then follow their propagation through the Earth's crust to the detector. We simulate the DM particle interactions in the detector and calculate the rate of the detector's response. We compare it to the measured rate and extract cross section limits.
The basic input parameters of our calculation are the composition of the target, the depth of the detector and the energy threshold of the experiment. We also show the value of the dark matter mass threshold $m_{TH}$, calculated for the standard and conservative parameter values given in (\ref{parameters}) and (\ref{range}). The parameters are summarized in Table \ref{t1} for the relevant experiments.

\begin{table}[htb] \label{t1}

\caption{The parameters of the experiments used for the extraction of cross section limits; $m_{TH}$ is the minimum mass of DM particle which can produce a detectable recoil, for the standard and conservative parameter choice. The energy threshold values $E^{nuc} _{TH}$ refer to the nuclear response threshold. This corresponds to the electron response threshold divided by the quenching factor of the target.} \label{t1}
\begin{center}
\begin{tabular}{|c|c|c|c|c|}
\hline

Experiment                   &     Target    &  Depth     & $E^{nuc} _{TH}$ & $m^{std}_{TH} (m^{cons}_{TH})$      \\  \hline \hline
CRESST, \cite{cresst}        &  Al$_2$O$_3$  &  1400 m    & 600 eV & 0.8 (1.1) GeV   \\ \hline
DAMA, \cite{dama}            &  NaI          &  1400 m    & 6  keV & 3.5 (5) GeV \\ \hline
ELEGANT, \cite{elegant}      &  NaI          &   442 m    & 10 keV & 5 (8) GeV   \\ \hline
COSME I, \cite{cosme}          &  Ge           &  263 m     & 6.5 keV & 5.5 (8) GeV        \\ \hline
CDMS, \cite{cdms}            &  Si, Ge       &  10.6 m    & 25 keV & 9.8 (16) GeV   \\ \hline

\end{tabular}
\end{center}
\end{table}

In the code, after generating particles we propagate those which satisfy $<{\hat n} \cdot {\vec v}>\leq 0$ through the crust. Given the effective mean free path in the crust (\ref{freepath}), the distance traveled between two collisions in a given medium is simulated as  
\beq \label{distance}
x=-\lambda _{eff} \ln [R]
\eeq
where R is a uniformly distributed random number, in the range $(0,1)$.
After simulating the distance a particle travels before scattering, we choose the nucleus $i$ it scatters from using the relative probability as in (\ref{probability}). 

We take the mass density of the crust to be $\rho =2.7$ g/cm$^3$. To explore the sensitivity of the result to the composition of the crust we consider two different compositions. First we approximate the crust as being composed of quartz, ${\rm SiO}_2$, which is the most common mineral on the Earth and is frequently the primary mineral, with $>98 \%$ fraction. 
Then we test the sensitivity of the result by using the average composition of the Earth's crust: Oxygen 46.5 $\%$, Silicon 28.9 $\%$, Aluminium 8.3 $\%$ and Iron 4.8 $\%$, where the percentage is the mass fraction of the given element. Our test computer runs showed that both compositions give the same result up to the first digit, so we used simpler composition for the computing time benefit. Since the DM exclusion window we obtain at the end of this section should be very easy to explore in a dedicated experiment, as we show later in the text, we do not aim to find precise values of the signal in the underground detector. 

When collisions reduce the DM velocity to less than the Earth's radial escape velocity, $v_{esc}=11$ km/s, DM is captured by the Earth and eventually thermalized. Collisions may also reverse the DM velocity in such a way that the DM particle leaves the surface of the Earth with negative velocity: effectively, the DM particle gets  reflected from the Earth. The majority of light DM particles wind up being reflected as is characteristic of diffuse media. The percentage of reflected particles proves not to depend on the cross section, as long as the mean free path is much smaller than the radius of the earth, but it does depend on DM particle mass. Light particles have a higher chance of scattering backward and therefore a higher percentage of them are reflected. The initial DM flux on Earth equals $2.4(1.2)~10^{6}~(1~{\rm GeV}/m_X)$ cm$^{-2}$s$^{-1}$, taking standard (conservative) parameter values. Table \ref{t2} shows the fraction of initial flux of DM particles on the Earth which are captured and thermalized for various mass values. The fraction is, up to a percent difference, independent of whether we make the standard or conservative parameter choice. 

For DM particles which are not scattered back to the atmosphere and which pass the depth of the detector before falling below the energy threshold of the detector, the scattering in the detector is simulated. 
\begin{table}[htb]

\caption{ The percentage of DM particles incident on Earth which are captured, when $\lambda _{int} << R_E$.} \label{t2}
\begin{center}
\begin{tabular}{|c|c|c|c|c|c|c|}
\hline

mass [GeV]            &   2       &  4     & 6       & 10       & 100       \\  \hline \hline

thermalized [$\%$]    &   21  & 30 & 36  & 46   & 94    \\ \hline

\end{tabular}
\end{center}
\end{table}
For composite targets we simulate collision with different nuclei with probabilities given as in (\ref{probability}). If the energy of the recoil is above $E_{TH}$, we accept the event and record the velocity of the particle which deposited the signal.
The spectral rate per (kg day keV) is then calculated as a sum of rates on the separate elements of the target, as
\beq \label{sum}
\frac{dR}{dE_R} (\alpha (t))=\sum_{i} \frac{f_i}{A_i~m_p}~\frac{\rho _X}{m_X} ~\frac{<v(\alpha (t))>_i}{\Delta E}~\sigma _{XA_i}
\eeq
where $f_i$ is the mass fraction of a given element in the target, $\rho _X$ is the local DM energy density, $\Delta E$ is the size of an energy bin of a given experiment and $<v(\alpha (t))>$ is calculated as in (\ref{<v>}). The signal in the detector falls exponentially with $\sigma _{XN}$ since the energy of DM at a given depth gets degraded as an exponential function of the cross section, see \cite{SG}. Therefore the limit on $\sigma _{XN}$ is insensitive to small changes in the rate in the detector coming from changes in $\rho _X$;  we adopt the commonly used value $\rho _X=0.3$ GeV cm$^{-3}$ for the local DM energy density. 

We emphasize here that the spectral rate is a function of the relative angle $\alpha (t)$ between the direction of the motion of LSR and the position of the  detector.
This angle changes during the day as
\beq \label{time}
\cos \alpha (t)=\cos \delta \cos \alpha '+ \sin \delta \sin \alpha ' \sin (\omega t +\phi _0)
\eeq 
where $\delta $ is the angle between the Earth's North pole and the motion of LSR; $\alpha '$ is the angle between the position of the detector and the North pole, and it has a value of ($90^{\circ}$ - geographical latitude).
The angle between the LSR motion and Earth's North pole is $\delta = 42^{\circ}$, so for an experiment around 45$^{\circ}$ latitude (as for Gran Sasso experiment), $\alpha '=45 ^{\circ}$. Therefore, in the course of a day, the angle between the detector and the LSR motion varies in the range approximately 0$^{\circ}$ to 90$^{\circ}$.

Fig. 3 shows the rate $R$ per (kg $\cdot $ day) as a function of time, (\ref{time}), 
\beq
R (\alpha (t))=\sum_{i=Al,O} \frac{f_i}{A_i~m_p}~\frac{\rho _X}{m_X} ~<v(\alpha (t))>_i~\sigma _{XA_i}
\eeq
calculated for the parameters of the CRESST experiment.
We choose $\phi _0$ so that time t=0 corresponds to the moment the detector starts to move away from the LSR axis. We see that for these masses the rate is a strong function of the angle of the position of the detector with respect to the motion of the LSR, which gives an interesting detection signature for detector locations such that this angle changes during the day.  

\vspace{3mm} \label{figdaydep}
\centerline{\epsfig{file=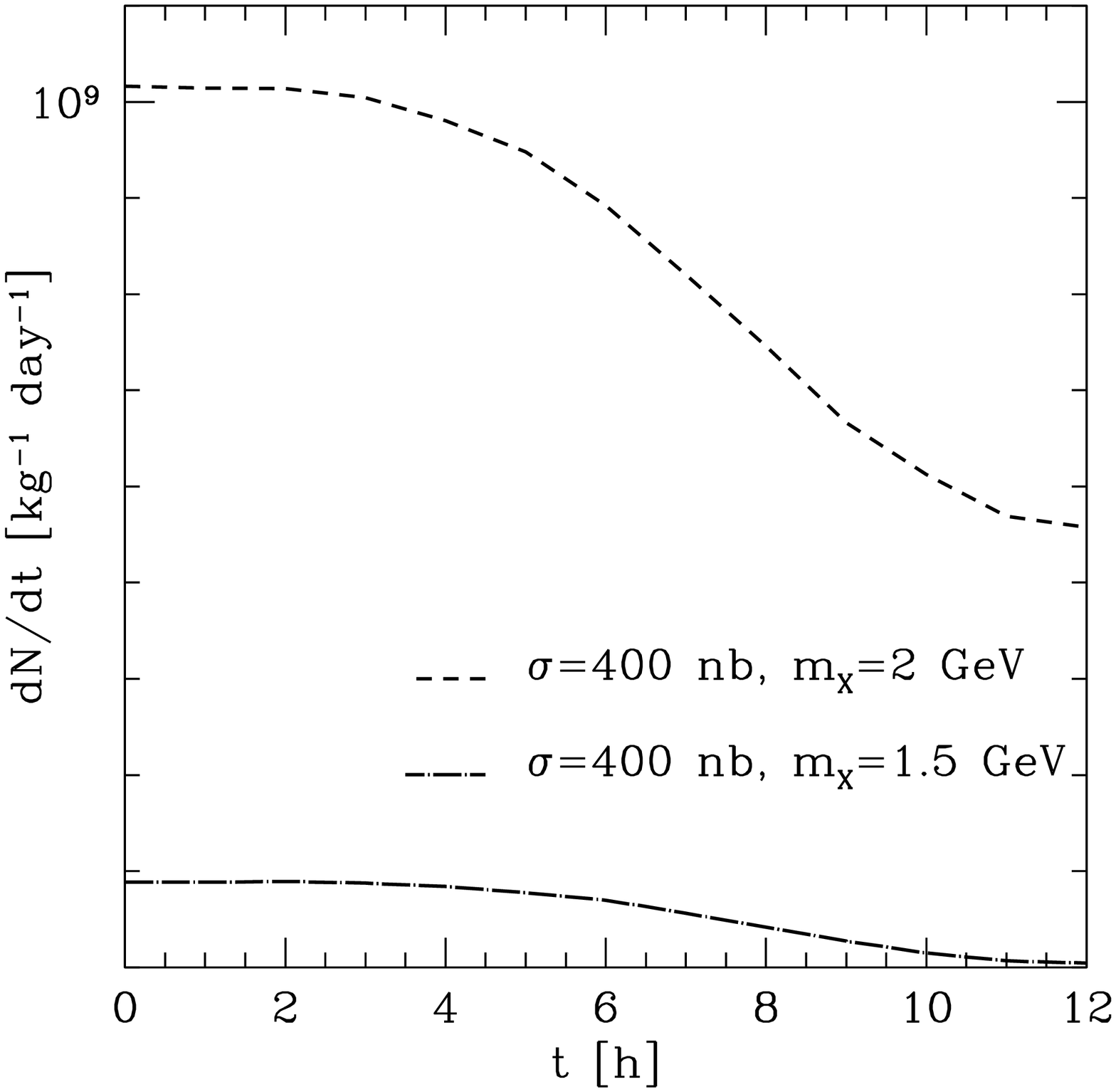,width=6cm}}
{\footnotesize\textbf{Figure 3:} The time dependence of the measured rate in underground detectors for m=2 GeV and m=1.5 GeV DM candidates.}
\vspace{3mm}

To extract our limits, we average the signal from the simulation $dR(t)/dE_R$ over one day: 
\beq
<dR/dE_R>=\frac {1}{T}~\int ^{T} _0 dR(t)/dE_R~dt.
\eeq
Since the shape of the spectral rate response is a function of $\sigma _{Xp}$ in our case (because the velocity distribution function at the depth of detector is a function of $\sigma _{Xp}$ due to the interactions in the crust) the extraction of cross section limits is more complicated than when the rate scales linearly with $\sigma _{Xp}$. In the region where the window is located, i.e. masses below 2 GeV, we perform the analysis based on the fit method used by the CRESST group, \cite{cresst}. The measured spectrum is fit with an empirical function called $B$. In our case $B$ is the sum of two falling exponentials and a constant term, since we expect the signal only in the few lowest energy bins. For the fit we use the maximum likelihood method with Poissonian statistics in each bin. The maximum likelihood of the best fit, $B_0$, is ${\emph L}_0$. We define the background function $B'$ as the difference between the best fit to the measured signal, $B_0$ and some hypothesised DM signal $S$: $B'=B_0-S$. Following the CRESST procedure, we set $B'$ to zero when $S$ exceeds $B_0$. When $\sigma _0$ is such that the simulated signal $S$ is below the measured signal $B_0$, $B'$ adds to the signal $S$, completing it to $B_0$ and the likelihood is unchanged. With increasing $\sigma _0$, when $S$ starts to exceed the function $B_0$, $B'$ becomes zero, and we calculate the likelihood using $S$ alone in such bins, leading to a new likelihood ${\emph L}$. Following the CRESST collaboration prescription, $\sigma_0$ excluded at $90 \%$ CL is taken to be the value of $\sigma_0$ giving $\ln{\emph L}-\ln{{\emph L}_0}=-1.28^2/2$ \cite{cresst}, since $10\%$ of the area of a normalized Gaussian distribution of width $\sigma$ is $1.28 \sigma $ above than the peak. We show the window obtained this way in Figure 2 and for the low mass range, in Figure 1. 

For masses higher than $2$ GeV we can use a simpler method, since this range of masses is already ruled out and our plot is only indicating from which experiments the constraints are coming. We calculate the response of the detector for different cross section values, and take the limit to be that value for which the simulated signal is below the experiment's  background. Fig 4 shows CRESST background together with the simulated response for DM particles with mass $m_X=2$ GeV and 10 GeV and various values of cross section. The limits obtained this way for different experiments are given in Figure 1.

\vspace{3mm} \label{signal}
\centerline{\epsfig{file=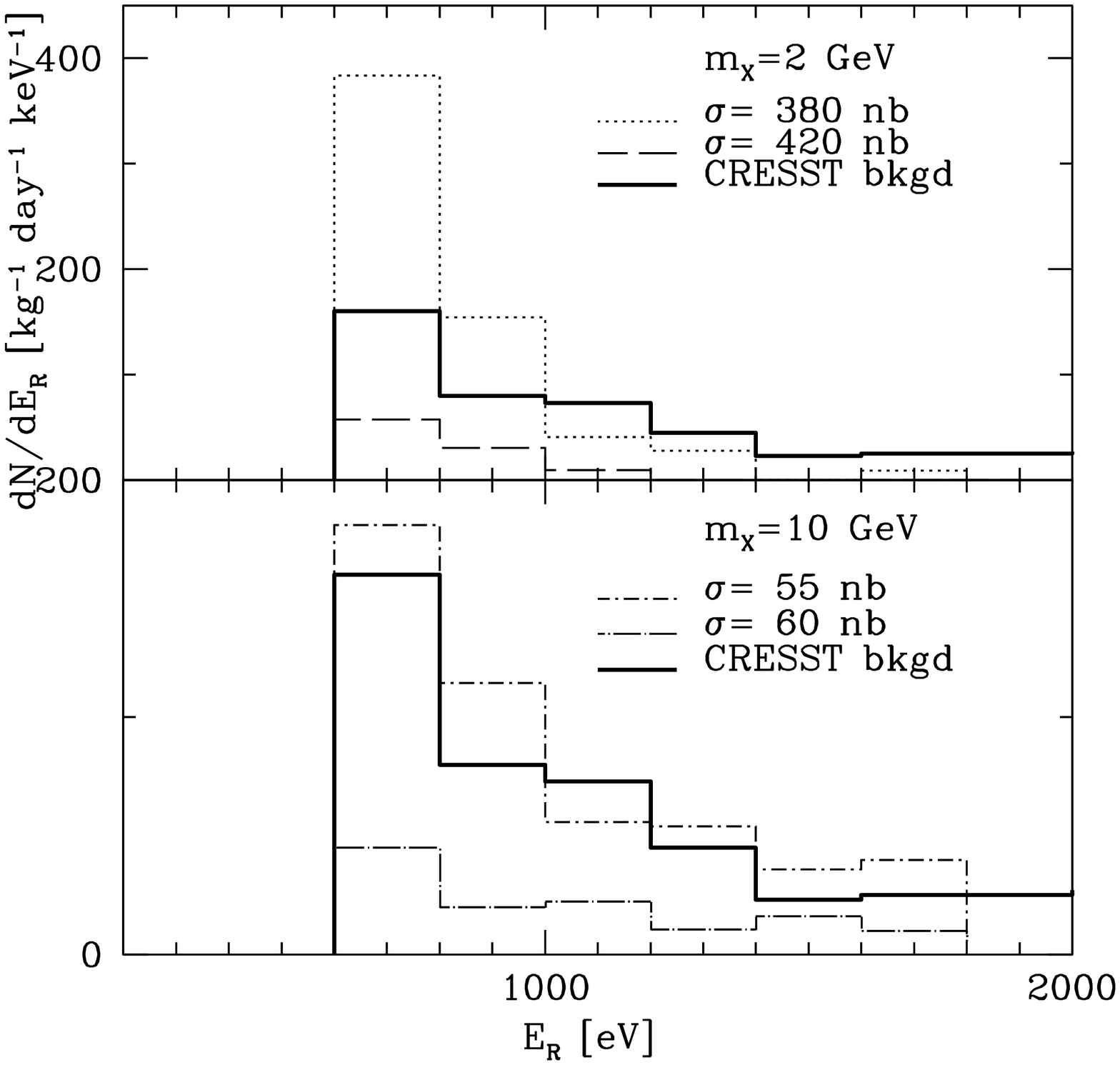,width=8cm}}
{\footnotesize\textbf{Figure 4:} The CRESST background and the simulated response of the detector for masses $m_X=2$ and $m_X=10$ GeV, and different values of spin independent cross sections $\sigma _{Xp}$.}
\vspace{3mm} 

The only dark matter detector sensitive to particles with mass $\lsi 4$ GeV is CRESST. Since it is the only experiment with threshold lower than the threshold of the balloon experiment by Rich et al., it extends the existing exclusion window for intermediate cross sections. For the CRESST experiment we perform the calculation using both standard and conservative parameters, because the size of the exclusion window is very sensitive to the value of mass threshold, and therefore to the parameter choice. For other underground experiments we use only standard parameters. In the mass range $5\lsi m\lsi 10$ GeV, the ELEGANT and COSME I experiments place the most stringent constraints on a DMIC scenario, since they are located in shallow sites a few hundred meters below the ground; see Table \ref{t1}. Other experiments sensitive in this mass range (e.g. IGEX, COSME II) are located in much deeper laboratories and therefore less suitable for DMIC limits. We therefore present limits from ELEGANT and COSME I, for masses 5 to 10 GeV. Masses grater than 10 GeV are above the threshold of the CDMS experiment and this experiment places the most stringent lower cross section limit due to having the smallest amount of shielding, being only 10.6 m under ground. Fig 1 shows the cross section limits these experiments imply, for masses $m\lsi 10^3$ GeV. 
\vspace{3mm}

\centerline{\epsfig{file=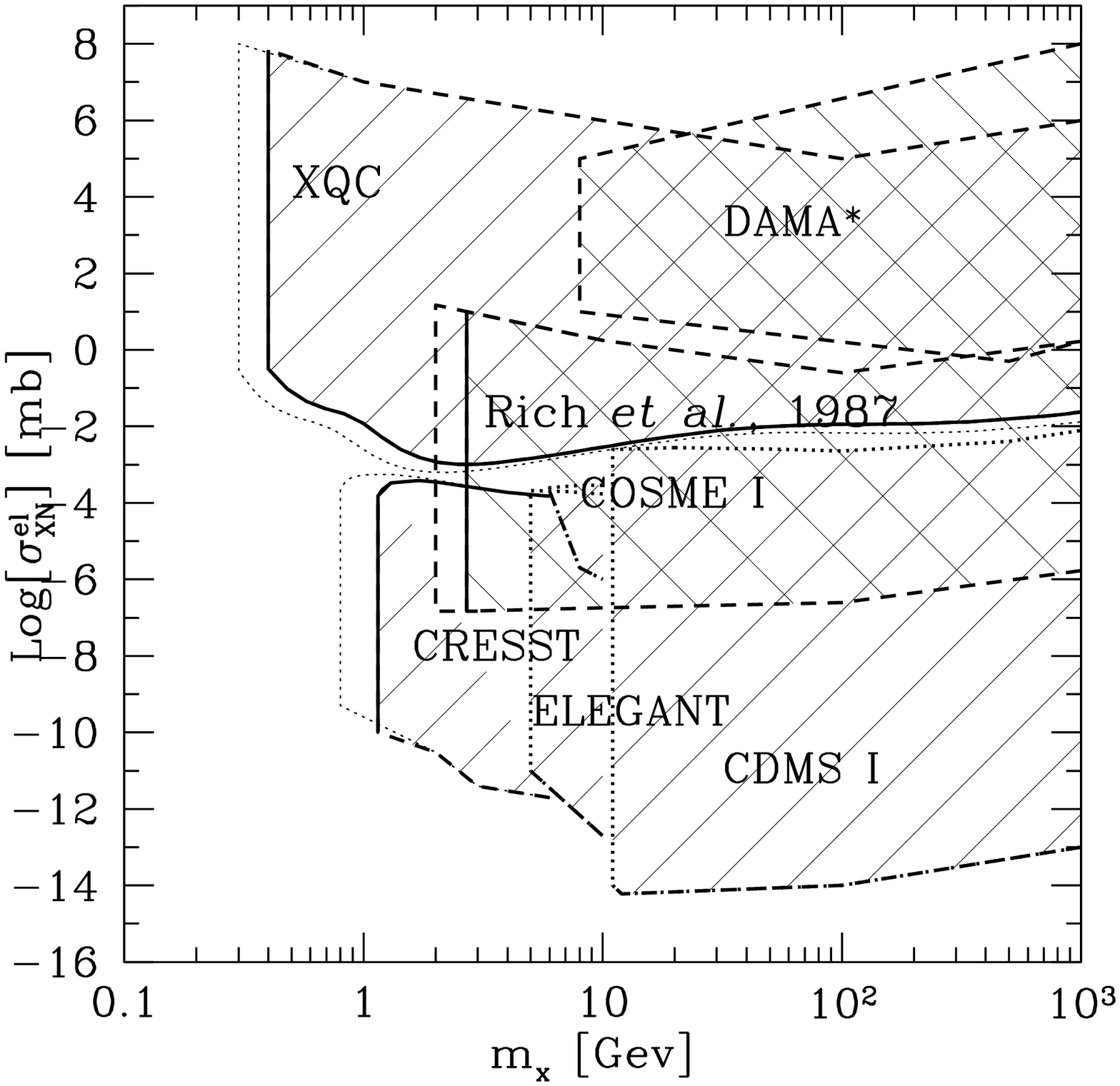,width=8cm}} 
{\footnotesize\textbf{Figure 1:} Overview of the exclusion limits for spin independent DM-nucleon elastic cross section coming from the direct detection experiments on Earth. The limits obtained by using the conservative parameters, as explained in the text, are plotted with a solid line; the dotted lines are obtained by using standard parameter values and the dashed lines show limits published by corresponding experiments or in the case of XQC, by Wandelt et al. \cite{SS}. The region labeled with DAMA* refers to the limits published in \cite{damastrong}.}
\vspace{3mm} 

\vspace{3mm} \label{figSI}
\centerline{\epsfig{file=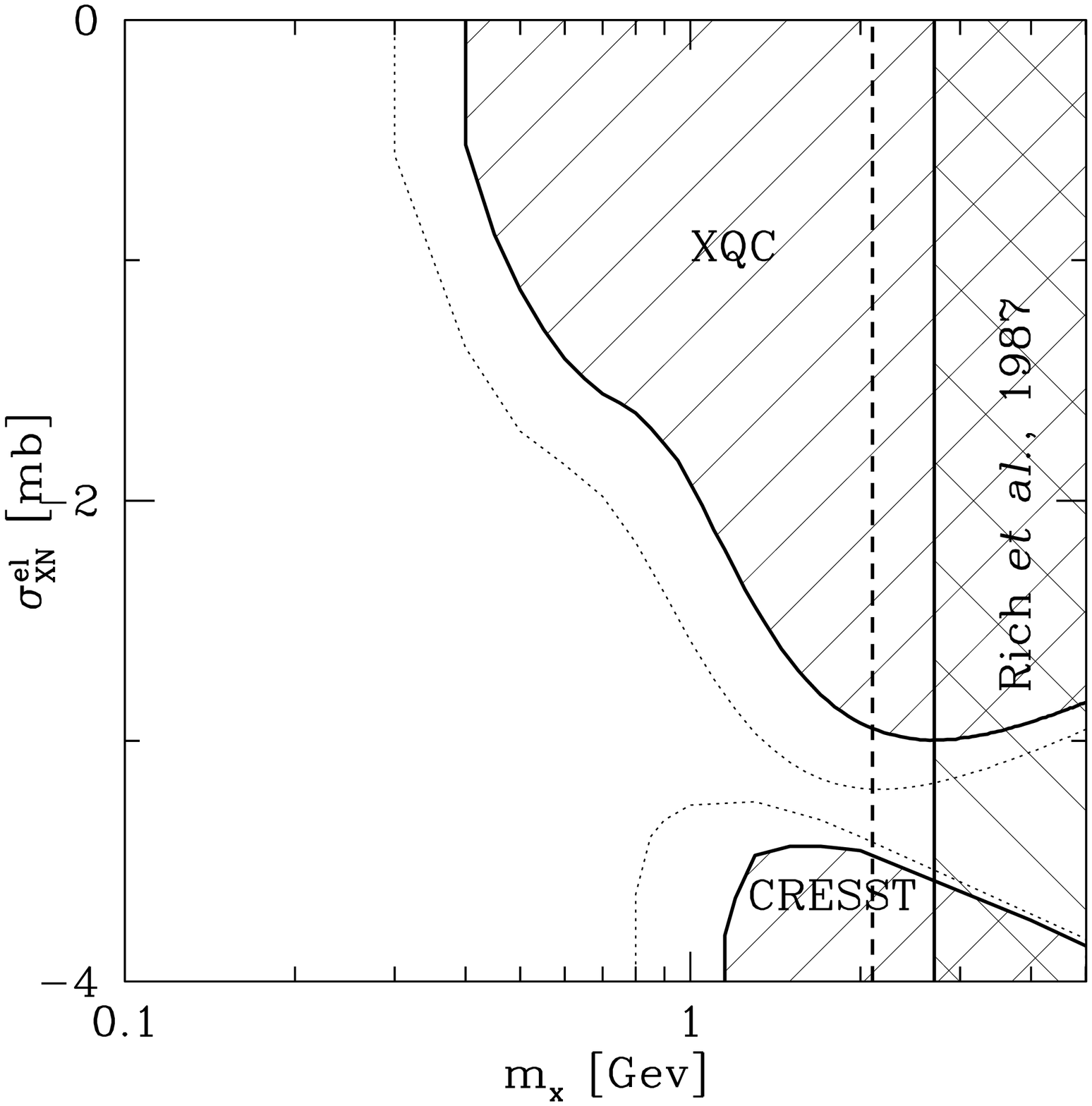,width=8cm}}
{\footnotesize\textbf{Figure 2:} The allowed window for $\sigma ^{el} _{XN}$ for a spin independent interaction. The region above the upper curve is excluded by the XQC measurements. The region below the lower curve is excluded by the underground CRESST experiment. The region $m\gsi 2.4$ GeV is excluded by the experiment of Rich et al..}
\vspace{3mm}

\section{Spin-Dependent limits} \label{SD}

In this section we address the case in which DM has a spin dependent interaction with ordinary matter. We consider first purely SD interaction and later we consider the case in which both interaction types are present. We focus on low masses which belong to the cross section window allowed by the experiment of Rich et al..
 
\vspace{3mm} \label{figSD}
\centerline{\epsfig{file=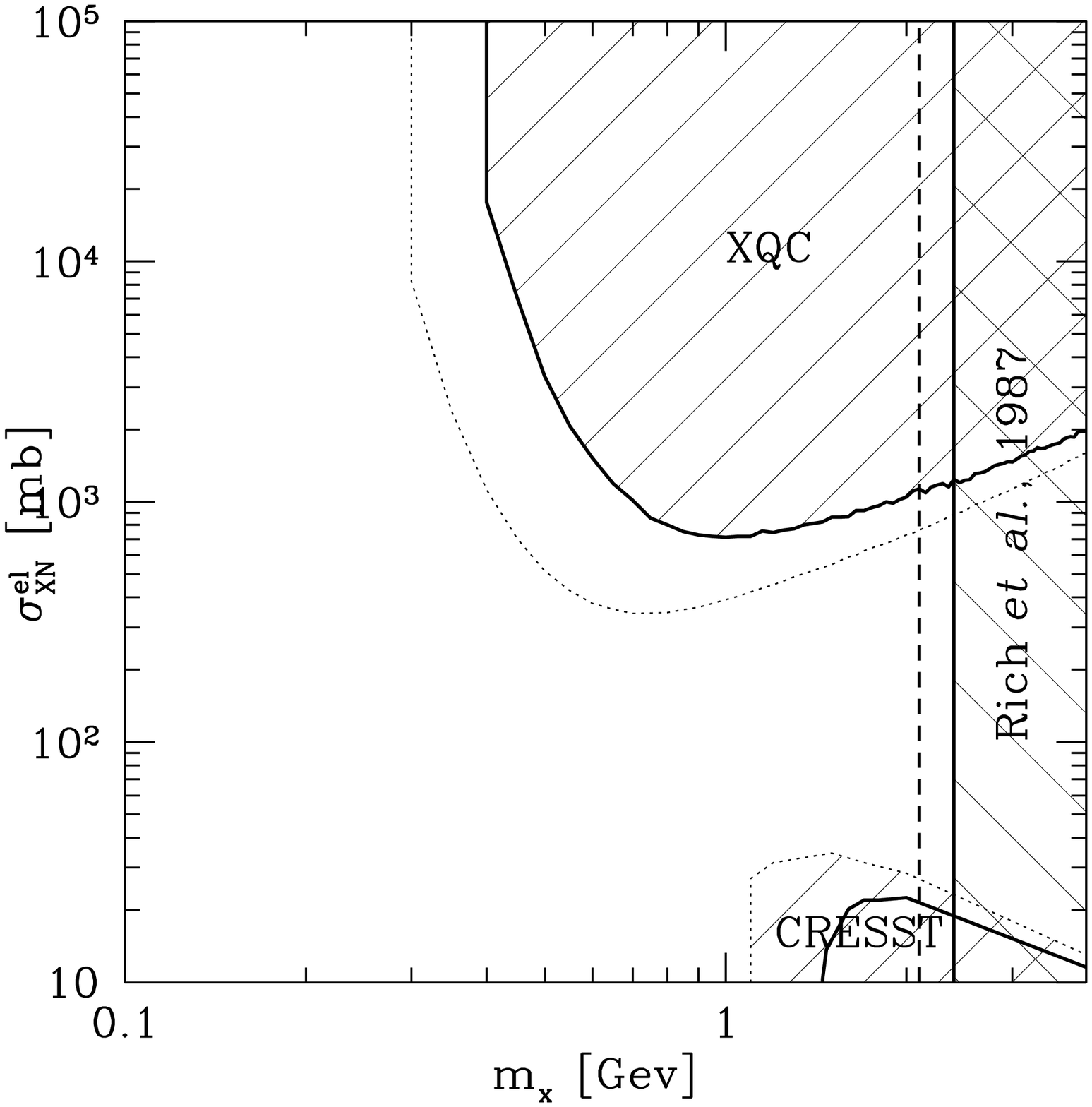,width=8cm}}
{\footnotesize\textbf{Figure 5:} The allowed Spin Dependent interaction for $(C_{Xp}/C_{Xn})^2=1$. The region above the upper curve is excluded by XQC measurements. The region below the lower curve is excluded by CRESST. The region $m\gsi 2.4$ GeV is excluded by the balloon experiment of Rich et al..}
\vspace{3mm}

If the DM has only a spin dependent interaction with ordinary matter, only the small fraction of the XQC target with nonzero spin is sensitive to DM detection. The nonzero spin nuclei in the target are: ${\rm Si}_{29}$ ($4.6~\% $ of natural Si), ${\rm Te}_{125}$ ($7~\% $) and ${\rm Hg}_{199}$, ($16.87~ \% $); their spin is due an unpaired neutron. We calculate the spin dependent cross section limits from the XQC experiment the same way as for the spin independent case, using the new composition of the target. The limiting value of the elastic cross section of DM with protons, $\sigma^{SD} _{Xp}$, is shown in Figure 5. Since the XQC target consists of n-type nuclei, the resulting cross section with protons is proportional to the $(C_{Xp}/C_{Xn})^2$ factor as explained in section II. In Figure 4 we use the value $(C_{Xp}/C_{Xn})^2=1$ which is the minimal value this ratio may have. We note that the maximal value of the ratio, based on the EMC measurements is $(C_{Xp}/C_{Xn})^2=500^2$ and it would shift the XQC limit by a factor $500^2$ up to higher cross sections (substantially extending the allowed window).

The spin sensitive element in the CRESST target is Al which has an unpaired proton in the natural isotope. We assume that the crust consists only of Al, since it is the most abundant target nucleus with non-zero spin. In this case the model dependence of the $C$ factor ratio drops out in normalizing the underground experiment to the proton cross section. 

\vspace{3mm} \label{figSISD}
\centerline{\epsfig{file=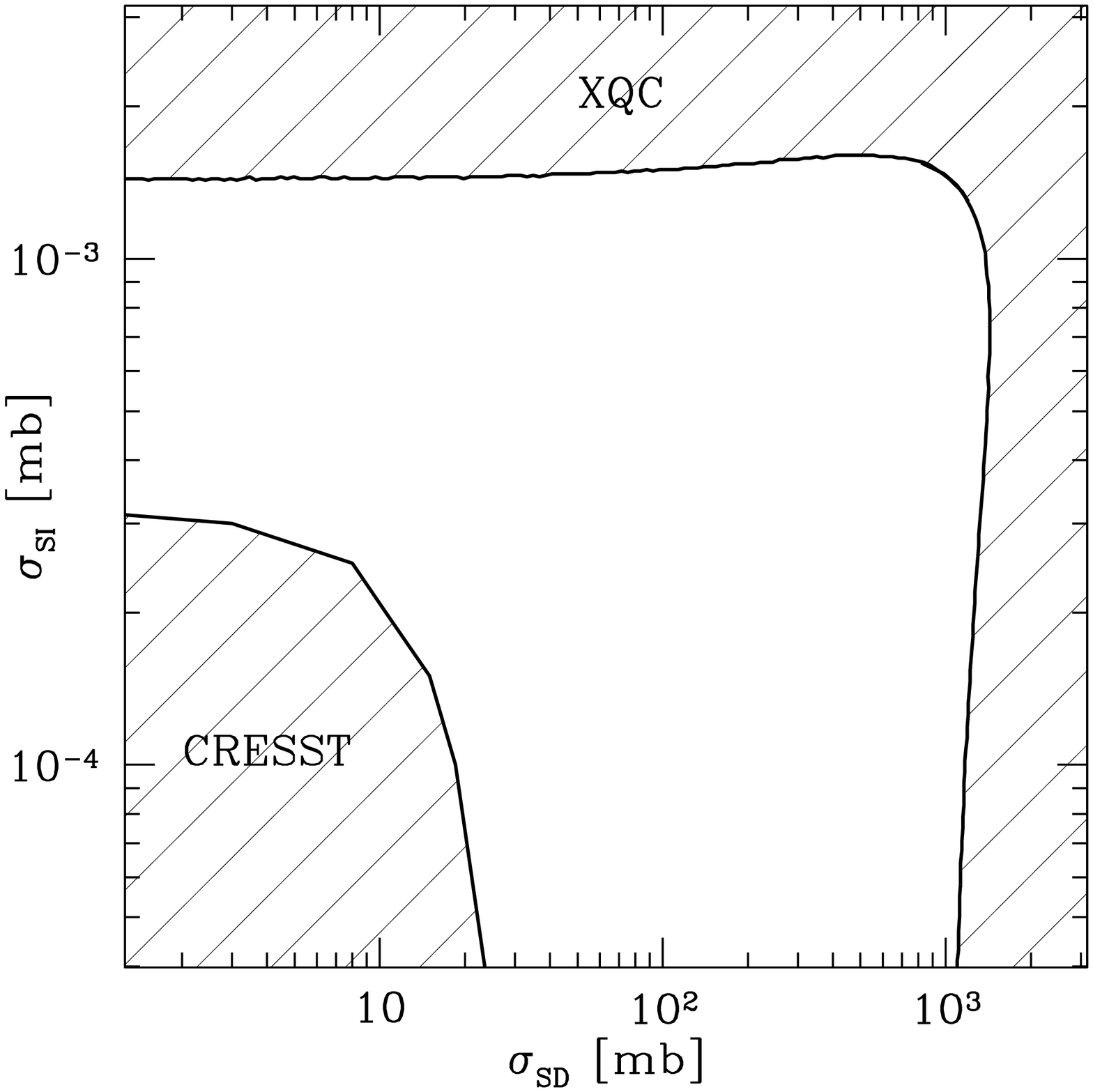,width=8cm}}
{\footnotesize\textbf{Figure 6:} $\sigma _{SI}$ vs $\sigma _{SD}$, for CRESST and XQC experiments, for mass $m_X=2$ GeV. The region between two curves is the allowed region.}
\vspace{3mm}

The window is extended when compared to the purely spin independent DM interaction, as shown in Fig. 5. This is mostly due to the fact that sensitive part of the XQC target is substantially reduced.

In Fig. 6, for mass $m_X=2$ GeV, we plot the $\sigma _{SI}$ vs $\sigma _{SD}$ limit, assuming both types of interactions are present. An interesting feature in the $\sigma _{SI}$ vs $\sigma _{SD}$  dependence is that, when the spin dependent and independent cross sections on the target nuclei are of comparable magnitude, screening between two types of targets allows cross sections to be higher for the same rate in the detector than in the case when only one type of interaction is present.

\section{Constraint on the fraction of DMIC}  \label{fraction}

We now turn the argument around and use the XQC data to place a constraint on the fraction of allowed DMIC as a function of its elastic cross section. We restrict consideration to values of the cross section which are small enough that we do not have to treat energy loss in the material surrounding the sensitive components of XQC. The maximal fraction DMIC allowed by XQC data $p=n^{MI} _{DM}/n^{tot} _{DM}$ can then be expressed as a function of cross section, using (\ref{rate})
\begin{eqnarray} \label{eqfraction}
p&=&\frac{N_S}{n_X~f~T} \{ N_{\rm Si}  <{\vec v}_{\rm Si}> \sigma_{\rm Si}    \\
&+& N_{\rm Hg} ( <{\vec v}_{\rm Hg}> \sigma_{\rm Hg}+<{\vec v}_{\rm Te}> \sigma_{\rm Te} )\} ^{-1} \nonumber
\end{eqnarray}
where all quantities are defined as before. 

The mass overburden of XQC can be approximated as \cite{mccam:correspondence}: $\lambda =10^{-4}~{\rm g/cm}^2$, for off-angle from the center of the field of the detector $\alpha =(0^o~{\rm to}~ 30^o)$; $\lambda =10~{\rm g/cm}^2$, for $\alpha =(30^o~{\rm to}~ 100^o )$; and  $\lambda =10^4~{\rm g/cm}^2$, for $\alpha \ge 100^o$. The center of the field of view points toward $l=90^o$, $b=60^o$ which makes an angle of $32^o$ with the detector position direction. Since DM particles are arriving only from above the detector, they will traverse either 10 g/cm$^3$ or $10^{-4}$ g/cm$^3$ overburden.  
\vspace{3mm}
\centerline{\epsfig{file=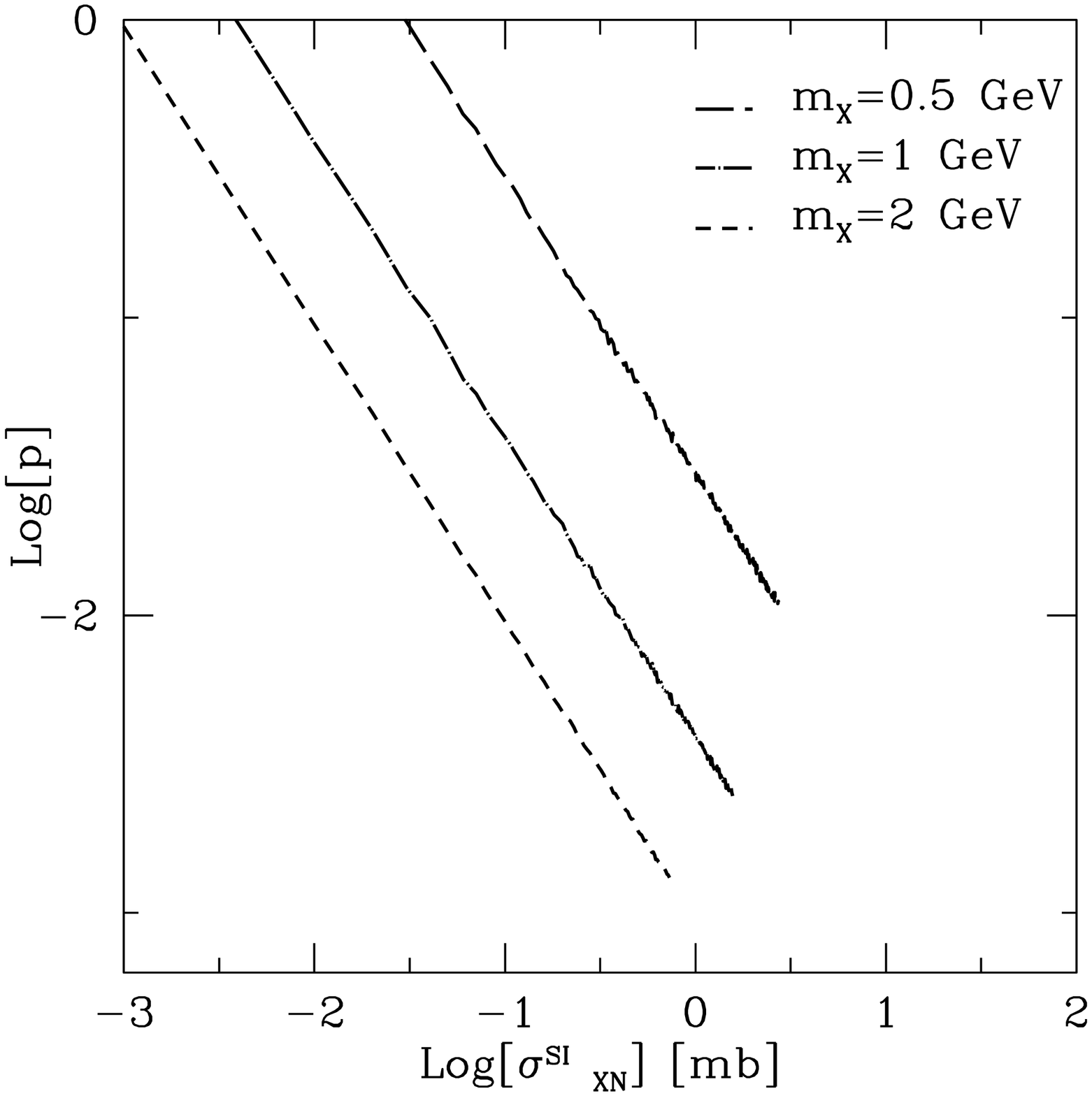,width=8cm}}
{\footnotesize \textbf{Figure 7}: The allowed fraction $p$ of DM candidate as a function of DM-nucleon cross section. For each mass, $p$ is calculated up to the values of cross sections for which the interaction in the mass overburden of the detector becomes important.}
\vspace{3mm}

For example, for values of cross section of about $0.7$ mb, $m=2$ GeV DM particles start to interact in the 10 g/cm$^3$ overburden, thus for cross sections above this value our simple approach which does not account for the real geometry of the detector, is not applicable anymore.
We therefore restrict our analysis to values of the cross section for which neglecting the interaction in the overburden is a good approximation. In this domain, the allowed fraction of DM falls linearly with increasing cross section, as can be seen in equation (\ref{eqfraction}) since $<{\vec v_{DM}}>$ remains almost constant and is given by the halo velocity distribution (\ref{veldistE}).
The results of the simulation are shown in Fig. 7, for a spin independent interaction cross section. 
An analysis valid for larger cross sections, which takes into account details of the geometry of the XQC detector, is in preparation \cite{Spergel}.

\section{Future Experiments}

The window for $m_X\lsi 2.4$ GeV in the DMIC cross section range could be explored in a dedicated experiment with a detector similar to the one used in the XQC experiment and performed on the ground. Here we calculate the spectral rate of DM interactions in such detector, in order to illustrate what the shape and magnitude of a signal would be. 
\vspace{3mm} \label{figftrexp}
\centerline{\epsfig{file=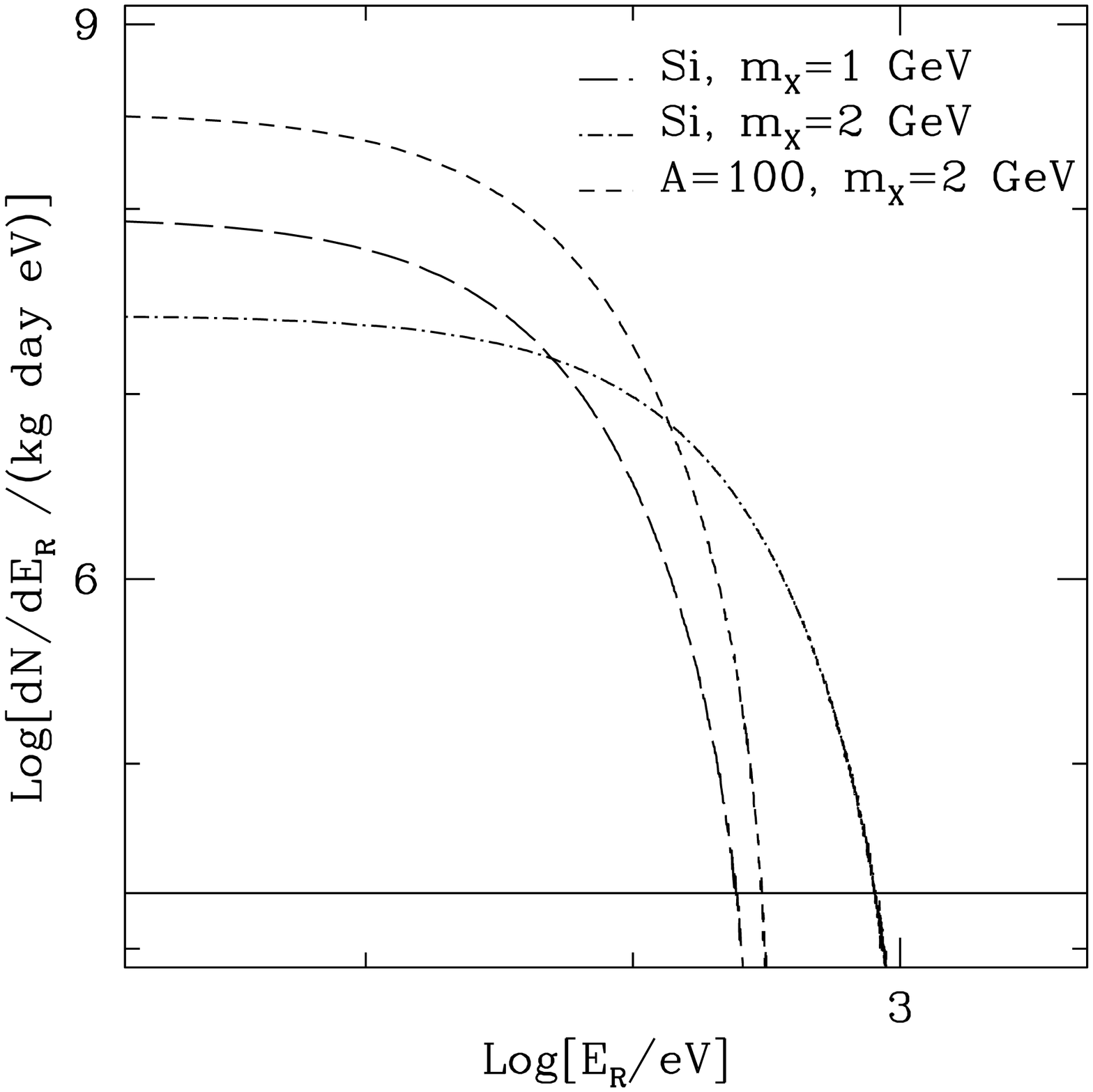,width=8cm}}
{\footnotesize \textbf{Figure 8}: The simulated minimum rate per (kg day eV) calculated with $\sigma _{Xp}=2~\mu$b, for a DM experiment on the ground, versus deposited energy $E_R$ in eV, for a SI target and for a target with mass number A=100. The solid line indicates maximal value of the cosmic ray muon background determined based on the total muon flux as is used in the text.}
\vspace{3mm}

In Fig. 8 we plot the rate per (kg$\cdot $day$\cdot $eV), for a Si detector and DM particle masses of $m_X=$ 1 and 2 GeV assuming a spin independent interaction. In the case of an exclusively spin dependent interaction, the signal would be smaller, approximately by a factor $f/A^2$, where {\it f} is the fraction of odd-nuclei in the target. The calculation is done for a position of a detector for which the signal would be the smallest. We assume a short experiment and do not perform averaging over time of a day because that would increase the signal.

The rate scales with cross section; the rate shown in Fig 8 is for $\sigma _{Xp}=2$ $\mu $b, the lower limit on the cross section window from the XQC experiment for $m=1$ GeV. Since the unshielded muon flux on the ground is of the order of $2~10^{2}~(m^2~s)^{-1}=2~10^3$ (cm$^2$ day$)^{-1}$, an experiment performed on the ground with an array of micro-calorimeter absorbers such as XQC whose target mass is $\approx 100$ g, should readily close this window or observe a striking signal.

\section{Conclusion}
In this paper we have determined the limits on dark matter in the low mass range ($m\lsi 10$ GeV) and with an intermediate cross section on nucleons based on the final XQC data and results of underground experiments with low mass threshold. We also updated previous limits taking into account newer halo velocity distribution. We found that there is an allowed window for DM mass $m\lsi 2.4$ GeV and cross section $\sigma \approx \mu$b. Curiously this window overlaps with the mass/cross section range expected for the H dibaryon making a possible DM candidate, \cite{fz}. We showed that it should be straightforward experimentally to explore the window. A signal due to a light DMIC would have strong daily variations depending on the detectors position with respect to the LSR motion and therefore provide strong signature.

We have benefited from discussions with D. McCammon, P. Huggins, T. Shutt, E. Sefusatti, S. Solganik, P. Steinhardt and L. Stodolsky.

\appendix
\section{} \label{appendix}
The probability $P (x+dx)$ that a particle will not scatter when propagating through a distance $x+dx$, equals the probability $P (x)$ that it does not scatter in the distance $x$, times the probability that it does not scatter from any type $i$ of target nuclei in the layer $dx$:
\beq 
P(x+dx)=P(x)\left(1-\sum \frac{dx}{\lambda_i} \right)\equiv P(x)\left(1- \frac{dx}{\lambda _{eff}} \right)
\eeq
By solving this differential equation one gets the probability that a particle will travel a distance $x$ in a given medium, without scattering,
\beq
P(x)=e^{-x/\lambda _{eff}}.
\eeq

The probability for scattering once and from a given nuclear species $i$ in the layer $(x,x+dx)$, is proportional to the product of probabilities that a particle will not scatter in distance $x$ and that it will scatter from species of type $i$ in $dx$:
\beq 
f_i(x)dx=e^{-x/ \lambda _{eff}}\frac{dx}{\lambda _i}.
\eeq

The probability that a particle scatters once from any species in a $dx$ layer is the sum of the single particle probabilities $\sum f_i(x)dx$, where
\beq
\int ^{\infty} _{0} \sum f_i(x)~dx=1.
\eeq 

In the simulation we want to generate the spectrum of distances a particle travels before scattering once from any of elements, using a set of uniformly distributed random numbers.
We can achieve this by equating the differential probability for scattering to that of a uniformly distributed random number,
\beq
\sum f_i(x)~dx=dR.
\eeq
After integrating 
\beq
\int ^x _0 \sum f_i(x)~dx=\int ^R _0 dR,
\eeq 
we get for the distribution of scattering distances $x$
\beq
x=-\lambda _{eff} ~\ln R.
\eeq
The relative frequency of scattering from a nucleus of type $i$, is then given by  
\beq 
\int ^{\infty} _{0} f_i(x)dx=\frac{\lambda _{eff}}{\lambda _i}=\frac {n_i\sigma _{XA_i}}{\sum  n_j\sigma _{XA_j}}.
\eeq 

%\appendix
\section{}

We assume the following function for the form factor, as explained in Section \ref{directdet},
\begin{equation}
F^2(q^2)=\exp ^{-\frac{1}{10}(qR)^2},
\end{equation}
where $q$ is momentum transfer and $R$ is the nuclear radius.
For a particle moving with a given velocity $v$, the mean free path to the next collision is obtained using the cross section $\sigma _{tot}$ which corresponds to $\sigma (q)$ integrated over the available momentum transfer range, from zero to $q_{max}$, where $q_{max}=2m_NE_{R,max}$ and $E_{R,max}=2\mu^2/m_A(v/c)^2$:
\begin{equation}
\sigma _{tot}=\sigma _0\frac{\int^{q_{max}} _0 F^2(q^2)dq^2}{\int^{q_{max}} _0 dq^2}.
\end{equation}
After a particle travels the distance calculated from the mean free path described above, the collision is simulated. The momentum transfer of a collision is determined based on the distribution given by the form factor function, as in the usual Monte Carlo method procedure
\begin{equation}
\int ^p _0 dp=\frac{\int ^q _0 F^2(q^2)dq^2}{\int ^{q_{max}} _0 F^2(q^2)dq^2},
\end{equation}
where $p$ is a uniformly distributed random number from $0$ to $1$.
Once the momentum transfer of the collision is determined, the recoil energy of the nucleus, $E_R$, and the scattering angle of the collision, $\theta _{CM}$, are uniquely determined. 

We repeat this procedure while following the propagation of a particle to the detector. If the particle reaches the detector we simulate the collision with target nuclei. For each collision in the target, the energy deposited in the detector $E_R$ is determined as above. For each particle $i$ the energy transfer determines the cross section with target nuclei as $\sigma_{XA_i} (E_R)=\sigma _{XA,0} F^2(E_R )$. The rate in the detector is found as in equation (\ref{sum}) with the only difference that in this case the sum runs over $\sum_{i} <v(\alpha (t)) \sigma _{XA_i}>_i$ instead of depending only on $v(\alpha (t))$.

%\bibliographystyle{unsrt}
%\bibliography{dmdet}

%\begin{thebibliography}{10}

%\end{thebibliography}

\end{document}